\documentclass[prd,aps,twocolumn,nofootinbib]{revtex4}

\newcommand{\ds}{\displaystyle}
\newcommand{\ov}{\overline}

\begin{document}

\title{Towards the robustness of the Affleck-Dine baryogenesis}

\author{Shinta Kasuya$^a$ and Masahiro Kawasaki$^b$}

\affiliation{
$^a$ Department of Information Science,
     Kanagawa University, Kanagawa 259-1293, Japan\\
$^b$ Institute for Cosmic Ray Research,
     University of Tokyo, Chiba 277-8582, Japan}

\date{June 9, 2006}

\begin{abstract}
We study the Affleck-Dine mechanism with various types of the K\"ahler potential, and 
investigate whether or not the Affleck-Dine field could acquire a large VEV as an initial
condition for successful baryogenesis. In addition to a negative Hubble-induced mass term, 
we find that large enough Hubble-induced A-terms could also develop the minimum at large 
amplitude of the field. Therefore, the Affleck-Dine mechanism works for broader classes
of the theories.
\end{abstract}


\maketitle

\setcounter{footnote}{1}
\renewcommand{\thefootnote}{\fnsymbol{footnote}}

\section{Introduction}
The observed universe mostly consists of matter, not antimatter. The asymmetry of the
baryons in the universe is known to be $\eta \sim 10^{-10}$, where $\eta$ is the 
baryon-to-entropy ratio, through the observations of light nuclei \cite{BBN} and cosmic
microwave background radiation \cite{WMAP}. It is one of the greatest issues of modern
cosmology how to create such an amount of the baryon asymmetry of the universe. 

In the context of supersymmetry (SUSY), a promising candidate of the baryogenesis is 
the Affleck-Dine mechanism \cite{AD,DRT}. It utilizes a scalar field carrying the baryon 
charge, which is called the Affleck-Dine field $\phi$. In particular in the minimal supersymmetric 
standard model, there are a lot of flat directions whose potential vanishes along those 
directions. Since the flat directions consist of squarks and/or sleptons, it is thus  
natural to regard them as the Affleck-Dine field.

During the inflationary stage, the Affleck-Dine field has a large vacuum expectation value (VEV). 
Well after inflation ends, it begins rotation in its potential when the Hubble parameter 
becomes the mass scale of the field, $H \sim m_\phi$. Since the baryon number 
(N\oe ther charge) is given by
\begin{equation}
Q = \int d^3 x \frac{1}{i}\left( \phi \dot{\phi}^* - \dot{\phi} \phi^* \right)
= \frac{1}{2} \int d^3 x \varphi^2 \dot{\theta},
\end{equation}
where $\phi=\varphi e^{i\theta}/\sqrt{2}$, the helical motion implies baryon number production.
In most cases, the Affleck-Dine field feels spatial instabilities, and deforms into Q balls 
\cite{qball}. From the decay or evaporation of the formed Q balls, quarks are produced 
afterwards, and we have a baryon asymmetry of the universe in usual sense.

The key ingredient for successful Affleck-Dine baryogensis is how to obtain a large VEV in 
the first place. During inflation, there appears a mass term due to SUSY breaking by the
finite energy density of the inflaton, which is called a Hubble-induced mass term. In supergravity
with the minimal K\"ahler potential, only a positive Hubble-induced mass term arises, which
does not make the field having a large VEV. Therefore, it is usually necessary to have 
nonrenormalizable terms in the K\"ahler potential to obtain a negative Hubble-induced mass 
term, $c_HH^2|\phi|^2$ with $c_H<0$.

In this paper, we investigate the cases when the field acquires a large VEV due to the negative Hubble-induced mass terms for some types of nonminimal K\"ahler potential. On the other hand,
we also consider the opposite situation that the Hubble-induced mass term is positive. Usually 
in this case, the Affleck-Dine field settles down to the origin of the potential, and cannot have a 
large VEV, which implies that the Affleck-Dine mechanism does not work.\footnote{
Successful scenarios exist for the case when the field amplitude becomes large 
during inflation, but the Hubble-induced mass term is positive after inflation. See 
Sect.~\ref{sec-positive} for the details. Similar aspects are investigated also in 
Ref.~\cite{McDonald}, especially in the context of D-term inflation.} 
The crucial observation, however, reveals that the potential will develop a (local or global) 
minimum at a large amplitude of the field due to Hubble-induced A-terms during and after 
inflation. 

The structure of the paper is as follows. We first review the Affleck-Dine mechanism in the 
usual case when the Hubble-induced mass term is negative in the next section. In Sect.III,
the origins of the Hubble-induced mass terms are investigated, and derive them for explicit 
examples of the K\"ahler potential in Sect.IV. In Sect.V, we study how the minima at large
amplitude of the field develop for large enough Hubble-induced A-terms while the 
Hubble-induced mass term is positive. Then, we consider the origins of the Hubble-induced 
A-terms, and obtain their explicit forms for $|I| \ll M_P$ in Sect.VI. Section VII is devoted for
investigating the situations that the large VEV is established during inflation, but the 
Hubble-induced mass term becomes positive after inflation. Finally, we summarize 
our conclusion in Sect.VIII. In Appendix, one can find the detail formulas of the Hubble-induced 
mass and A-terms for various K\"ahler potentials.

\section{Affleck-Dine mechanism due to a negative Hubble-induced mass term}
The potential of the flat direction vanishes only in the SUSY exact limit, and lifted by
SUSY breaking effects and nonrenormalizable operators. The general form of the potential
reads as
\begin{eqnarray}
\label{pot}
V(\phi) &  =  & m_\phi^2 |\phi|^2 + \left( A\frac{\phi^p}{pM_P^{p-3}} + h.c. \right) \nonumber \\
& +&  c_H H^2 |\phi|^2 + \left( a_H H \frac{\phi^q}{qM_P^{q-3}} + h.c. \right) \nonumber \\
& + & \lambda^2 \frac{|\phi|^{2(n-1)}}{M_P^{2(n-3)}}.
\end{eqnarray}
The first line represents the effect of (usual) SUSY breaking, while there are Hubble-induced 
mass and A terms in the second line due to finite energy density of the inflaton. 
Here $m_\phi \sim O$(TeV), $A \sim O(m_{3/2})$, $c_H \sim O(1)$, and $a_H \sim O(1)$. 
The last line comes from the nonrenormalizable superpotential of the form, 
$W(\phi)=\lambda \phi^n/nM_P^{n-3}$. In general, $p$, $q$, and $n$ could be different, but
$p=q=n$ in most cases, so we only treat this case hereafter otherwise mentioned.

Since, in order for the Affleck-Dine field to have a large VEV during inflation, the Hubble parameter is necessarily larger than $m_\phi$, the Hubble-induced terms dominate over the 
terms due to (hidden sector) SUSY breaking at that epoch. Thus, the first line of Eq.(\ref{pot}) 
is safely neglected when we consider the dynamics of the flat direction during inflation.

Let us first briefly remind the reader of the usual scenario of the Affleck-Dine mechanism. 
During inflation, the flat direction settles down in the minimum of the potential, which is
determined by the balance of the nonrenormalizable term (the third line of Eq.(\ref{pot})) and
the {\it negetive} Hubble-induced mass term (the first term in the second line of Eq.(\ref{pot})
with $c_H < 0$). Therefore, the amplitude of the minimum is estimated as 
$\varphi_{\rm min} \sim (HM_P^{n-3})^{1/(n-2)}$ where $c_H$, $\lambda \sim O(1)$ are assumed. 
After inflation when the Hubble parameter decreases as large as the mass of the flat direction, 
$H \sim m_\phi$, this minimum disappears and the flat direction begins moving towards the 
origin, the only (global) minimum. At the same time, the Hubble-induced and (hidden sector) 
SUSY breaking A terms become comparable. Since the Hubble parameter becomes also as 
large as the mass scale of the phase direction, 
$m_\theta \sim (A\varphi_{\rm min}^{n-2}/M_P^{n-3})^{1/2}$, the field feels torque due to the 
difference of the minima in the phase direction, and begins helical motion in the potential. 
This is the (dynamical) origin of the CP violation, one of the Sakharov's three conditions for 
baryogenesis. Thus, at the onset of oscillation in the potential, the baryon number density is 
estimated as, 
\begin{equation}
\label{baryon}
n_B \sim \frac{A}{m_\phi} \frac{\varphi_{\rm min}^n}{M_P^{n-3}} 
\sim  \left(\frac{m_\phi}{M_P}\right)^{\frac{n}{n-2}} M_P^3, 
\end{equation}
for $O(1)$ difference of the potential minima in the phase direction due to the usual and 
Hubble-induced A terms, and $A \sim m_\phi$ is used. In this scenario, the key is having a 
negative Hubble-induced mass term in order for the field to acquire a large VEV during inflation.

\section{Origin of the Hubble-induced mass terms}
In the supergravity the scalar potential is written in terms of superpotential, 
$W$, and K\"ahler potential, $K$, as
\begin{eqnarray}
&  & V=e^{K(\Phi, \Phi^{\dagger})/M_P^2}\Big[ \big( D_{\Phi_i} W(\Phi) \big)
K^{\Phi_i \ov{\Phi}_j} \big(D_{\ov{\Phi}_j} W^*(\Phi^{\dagger}) \big)  \nonumber \\
& & \qquad\qquad\qquad - \frac{3}{M_P^2} \left| W(\Phi) \right|^2 \Big] + ({\rm D-terms}),
\end{eqnarray}
where $\Phi$ denotes the scalar field in general, the subscript means the derivative with respect to the field, $F_\Phi \equiv D_{\Phi} W = W_{\Phi} + K_{\Phi} W/M_P^2$, and 
$K^{\Phi_i \ov{\Phi}_j}$ is the inverse matrix of $K_{\Phi_i \ov{\Phi}_j}$. Hereafter, 
we neglect the contribution from the D-term. In our argument, we consider only the flat 
direction $\phi$ and the inflaton $I$ with $W=W(\phi)+W(I)$.

In general, the Hubble-induced mass term arises from (a) the exponential prefactor as \cite{DRT}
\begin{equation}
\label{Hmass1}
e^{K(\phi,\phi^\dagger)/M_P^2} V(I),
\end{equation}
(b) cross terms in the K\"ahler derivative between the flat direction K\"ahler potential and
inflaton superpotential as \cite{DRT}
\begin{equation}
\label{Hmass2}
K_\phi K^{\phi\bar{\phi}}K_{\bar{\phi}} \frac{\left| W(I) \right|^2}{M_P^4},
\end{equation}
(c) K\"ahler potential couplings between the inflaton and the flat direction as \cite{DRT}
\begin{equation}
\label{Hmass3}
K_\phi K^{\phi\bar{I}} D_{\bar{I}} W^*(I^\dagger) \frac{W(I)}{M_P^2} +h.c.,
\end{equation}
and (d) K\"ahler derivative of the inflaton with nonminimal K\"ahler potential as
\begin{equation}
\big( D_I W(I) \big) K^{I \bar{I}} \big(D_{\bar{I}} W^*(I^{\dagger}) \big).
\end{equation}

During inflation the scalar potential is dominated by the energy of inflaton. We can thus write 
the effective potential of the inflaton as
\begin{eqnarray}
& & V(I) \simeq e^{K(I,I^{\dagger})/M_P^2}  \Big[ \big( D_I W(I) \big)
K^{I \bar{I}} \big(D_{\bar{I}} W^*(I^{\dagger}) \big)  \nonumber \\
& & \qquad\qquad\qquad\qquad - \frac{3}{M_P^2} \left| W(I) \right|^2 \Big].
\end{eqnarray}
In order to have positive potential energy, the first term in the parenthesis dominates:
\begin{equation}
|D_I W(I)| \gtrsim \frac{|W(I)|}{M_P}.
\end{equation}
Since the total energy density is dominated by the inflaton, we can relate it to the Hubble 
parameter as $V(I) \simeq 3 H^2 M_P^2$. In the inflaton oscillation dominated era after inflation,
the same formula is applicable if one regards $I$ as its amplitude. For $|I| \sim M_P$, we have
\begin{eqnarray}
D_I W(I) &\sim &HM_P, \\
W(I) & \lesssim & H M_P^2.
\end{eqnarray}
All of the contributions (a) $-$ (d) create the Hubble-induced mass term. However, the negative 
Hubble-induced mass terms should exist well after inflation until $H \sim m_\phi$, so we must 
thus seek for the case with $|I| \ll M_P$, even when $|I| \sim M_P$ during inflation. It is then 
necessary to equip nonminimal K\"ahler potential, because the minimal K\"ahler potential always 
results in a positive Hubble-induced mass term, as we will see in the next section. In this case, 
we have $K^{I\bar{I}}\simeq 1$, $|F_I| \simeq HM_P$, and $|W(I)| \ll HM_P^2$.  In the next 
section, we consider five types of K\"ahler potential which are typical examples, and see if it 
could result in negative Hubble-induced mass terms.

\section{Hubble-induced mass terms for $|I| \ll M_P$}

Now we consider if the Hubble-induced mass terms become positive or negative. We take
the following five (the minimal and four nonminimal) K\"ahler potentials as typical examples:
\begin{eqnarray}
K_m & = & \phi^\dagger \phi + I^\dagger I, \\
\delta K_1 & = & \frac{a}{M_P^2} \phi^\dagger \phi  I^\dagger I, \\
\delta K_2 & = & \frac{b}{2M_P} I^\dagger \phi \phi + h.c., \\
\delta K_3 & = & \frac{c}{4M_P^2} I^\dagger I^\dagger \phi \phi + h.c., \\
\delta K_4 & = & \frac{d}{M_P} I \phi^\dagger \phi + h.c.
\end{eqnarray}

For the minimal K\"ahler potential, only cases (a) and (b) are nonzero. As is well known,
in this case, the Hubble-induced mass term has positive coefficient:
\begin{equation}
c_H = 3 + \left(\frac{e^{K(I,I^\dagger)} |F_I|^2}{V(I)} -1\right) \simeq 3,
\end{equation}
where the last equality holds for $|I| \ll M_P$.

Therefore, nonminimal K\"ahler potential should be sought for obtaining negative Hubble-induced
mass terms. In each case we consider, we obtain the Hubble-induced mass term 
$c_H H^2|\phi|^2$ with
\begin{equation}
\label{Hmass}
c_H \simeq \left\{ \begin{array}{lcl} 
3(1-a) & & {\rm for} \ K_m + \delta K_1, \\[2mm] 
3(1+b^2) & &{\rm for} \ K_m + \delta K_2, \\[2mm]
3 & &{\rm for} \ K_m + \delta K_3, \\[2mm]
3(1+d^2) & & {\rm for} \ K_m + \delta K_4, \\
\end{array}  \right.
\end{equation}
for $|I| \ll M_P$.
The only possibility for a negative Hubble-induced mass term is introducing $\delta K_1$
with $a >1$.

If this is the only way for getting large VEVs during and after inflation, one may not seem
it very natural to have a successful Affleck-Dine mechanism. However, we show below that
large enough A-terms could lead the field to acquire large VEVs.

\section{Large VEV by Hubble-induced A-terms}
In this section, we describe how the effective potential acquires the minima at the large VEV
due to Hubble-induced A terms, even if the Hubble-induced mass term is positive. Considering 
only the second and third lines of Eq.(\ref{pot}), and rewriting as 
$\phi = \varphi \, e^{i\theta}/\sqrt{2}$, we have the potential of the form
\begin{eqnarray}
V(\varphi) & = & \frac{1}{2} c_H H^2 \varphi^2 
+  \lambda^2 \frac{\varphi^{2(n-1)}}{2^{n-1}M_P^{2(n-3)}} \nonumber \\
& & + \ a_H H \frac{\varphi^n}{2^{\frac{n}{2}-1}nM_P^{n-3}}\cos(n\theta).
\end{eqnarray}
For our purpose to obtain the minimum at large VEV, it is sufficient to set $\cos(n\theta)=-1$,
and consider only the particular radial direction with $n\theta=\pi$. It is then obvious that the 
$\varphi$ develops another minimum at $\varphi_{\rm min} \sim (H M_P^{n-3})^{1/(n-2)}$, 
provided that the following condition is met:
\begin{equation}
a_H^2 >  4(n-1) \lambda^2 c_H.
\end{equation}
Since the curvature at this minimum is of order $H^2$, the field rapidly settles down there 
during inflation. One might worry when this minimum is a local minimum. However, the transition
rate is approximately $P \sim \exp(-M_P^4/V(\phi)) \ll 1$ unless the dip and hill are extremely 
degenerate \cite{Linde}. Of course, one can set a little more severe condition 
$a_H^2 > n^2\lambda^2 c_H$, to make the dip as a global minimum. In any case, chaotic 
condition in the early inflationary stage will make the Affleck-Dine field settle into the minimum
at a large VEV with of order $O(1)$ probability.

The evolution of the field is very similar to that in the case of the negative Hubble-induced mass
term, since the field value of the newly developed minimum is almost the same if the parameters
such as $a_H$, $c_H$ and $\lambda$ are of order unity: 
$\varphi_{A, {\rm min}} \sim (H M_P^{n-3})^{1/(n-2)}$. After the field stuck into the 
minimum during inflation, it will stay there until $H\sim m_\phi$ when the Hubble-induced 
A term is overcome by the usual A-terms due to SUSY breaking by hidden sector. Thus, the 
field starts oscillation around the origin, and simultaneously feels torque to move along the 
phase direction. Since the field value and the power of the torque at the onset of the oscillation 
is the same as in the case of negative Hubble-induced mass term, the produced baryon number 
at that time is estimated as, for $O(1)$ difference in the phases,
\begin{equation}
n_B \sim \frac{A}{m_\phi} \frac{\varphi_{A,{\rm min}}^n}{M_P^{n-3}} 
\sim \left(\frac{m_\phi}{M_P}\right)^{\frac{n}{n-2}} M_P^3, 
\end{equation}
which is the same order of magnitude as Eq.(\ref{baryon}). Thus, the following evolution of the 
field should be similar, and hence we obtain almost the same amount of the baryon asymmetry 
of the universe.

\section{Hubble-induced A terms}

The Hubble-induced A terms arise, in general, from \cite{DRT} (i) cross terms in the 
K\"ahler derivative between the derivative of the flat direction superpotential and inflaton 
superpotential as
\begin{equation}
W_\phi K^{\phi\bar{\phi}}K_{\bar{\phi}}\frac{W^*(I^\dagger)}{M_P^2} + h.c.,
\end{equation}
(ii) cross terms between the flat direction superpotential and inflaton K\"ahler derivative as
\begin{equation}
K_I \frac{W(\phi)}{M_P^2} K^{I\bar{I}} \left( D_{\bar{I}}W^*(I^\dagger) \right) + h.c.,
\end{equation}
(iii) cross terms between the flat direction and inflaton superpotential as
\begin{equation}
- \frac{3}{M_P^2} W^*(I^\dagger) W(\phi) + h.c.,
\end{equation}
and (iv) K\"ahler potential couplings between the flat direction and inflaton as
\begin{equation}
W_\phi K^{\phi\bar{I}} \left( D_{\bar{I}}W^*(I^\dagger) \right) + h.c.
\end{equation}
For $|I| \ll M_P$, the minimal K\"ahler potential leads only to vanishing Hubble-induced A-terms,
so nonminimal ones are necessarily required, not only for developing minima at large VEV but
for obtaining the dynamical CP violation. The only nonvanishing Hubble-induced A terms appear
for $\delta K_2$ and $\delta K_4$ among which we considered:
\begin{eqnarray}
- b W_\phi \, \phi^\dagger H + h.c. & \qquad & {\rm for} \ \delta K_2, \\
- d \, W(\phi) H + h.c. & \qquad & {\rm for} \ \delta K_4. 
\end{eqnarray}
Although the A terms of the case with $\delta K_2$ look a bit weird, the abilities to have minima
at large amplitude and CP violation is the same. The only difference is that minima in the phase 
direction are fewer by two.

\section{Evolution in the positive Hubble-induced mass terms after inflation}
\label{sec-positive}
During inflation with $|I| \sim M_P$, there appear a lot of terms which destabilize the field from 
the origin, since $|I|/M_P \sim 1$ and/or $|W(I)|/M_P^2 \sim H$. In addition to A terms like 
$a_H H \phi^n/(n M^{n-3}) + h.c.$, another type of A terms of the form 
$\tilde{a}_H H^2 \phi \phi +h.c.$ for $|I| \sim M_P$ arise for some types of the K\"ahler potential.
These terms could develop the minima at large VEV if $\tilde{a}_H > c_H/4$.
(Complete lists of the Hubble-induced mass and A terms are shown in Appendix.) 
However, all of the terms other than those shown in the previous sections disappear
after inflation when $|I|\ll M_P$.\footnote{
If we consider non-additive superpotential such as $W(\phi,I)=(\phi/M_P)^sW(I)$, or alternatively
adopting a nonminimal K\"ahler potential of the form $f\phi^s/M_P^{s-2} +h.c.$, there could be 
A terms like $f H^2 \phi^s/M_P^{s-2} +h.c.$, which does not vanish for $|I| \ll M_P$.}
Then one may ask whether or not the Affleck-Dine mechanism 
works when the negative Hubble-induced mass and A terms vanishes after inflation.  We 
investigate the answer to it in this section. 

During inflation the Affleck-Dine field has a large VEV, 
$\varphi_{\rm inf} \sim (H_{inf} M_P^{n-3})^{1/(n-2)}$, due either to the negative Hubble-induced mass
or A terms. Thus, it will be the initial amplitude for the following evolution in the potential 
$V=c_H H^2 |\phi|^2$ with $c_H >0$. The evolution of the flat direction obeys
\begin{equation}
\ddot{\phi}+3H\dot{\phi}+c_H H^2 \phi =0.
\end{equation}
The amplitude of the flat direction decreases as $\varphi \propto H^{1/2}$ while oscillating 
around the origin for $c_H > 9/16$.  This is realized for the minimal K\"ahler 
potential which always exists,\footnote{We do dot consider theories such as no-scale supergravity.} after inflation with $c_H \simeq 3$ when $|I|\ll M_P$. Since the amplitude of 
the Affleck-Dine field is $\varphi \sim (H/H_{inf})^{1/2} (H_{inf} M_P^{n-3})^{1/(n-2)}$, the mass 
scale of the phase direction is given by 
\begin{equation}
m_\theta^2 \sim \frac{A \varphi^{n-2}}{M_P^{n-3}}
\sim  \frac{A}{H_{\rm inf}}  \left( \frac{H}{H_{\rm inf}}\right)^{\frac{n-6}{2}} H^2.
\end{equation}

The field starts moving towards the phase direction when $H \sim m_\theta$.  For $n=4$, 
it occurs when $H \sim m_\phi$. Thus the baryon number density can be estimated as
\begin{equation}
n_B \sim \frac{A}{m_\phi} \frac{\varphi_{A,{\rm min}}^4}{M_P} 
\sim \left(\frac{m_\phi}{M_P}\right)^2 M_P^3, 
\end{equation}
which is the same amount as in Eq.(\ref{baryon}). For $n \geq 6$, the mass scale is always 
much smaller than the Hubble parameter, $m_\theta^2 \ll H^2$, and the field only slow-rolls 
in the phase direction. Thus, the produce baryon number should be suppressed.
Since the baryon number evolves as 
\begin{equation}
 n_B a^3 \propto \int \! dt \, \varphi^n a^3 \propto \int \! dt \ t^{-\frac{n-4}{2}},
\end{equation}
for $H > m_\phi$, most of the baryon number is created at the beginning of oscillation just
after inflation. At that time, the baryon number is given by
\begin{equation}
n_{B,{\rm inf}} \sim \frac{A}{H_{\rm inf}} \frac{\varphi_{\rm inf}^n}{M_P^{n-3}} 
\sim \left(\frac{m_\phi}{H_{\rm inf}}\right)^{-\frac{2}{n-2}}
 \left(\frac{m_\phi}{M_P}\right)^{\frac{n}{n-2}} M_P^3, 
\end{equation}
In order to compare to the amount in the usual scenario (Eq.(\ref{baryon})), it should be
estimated at $H\sim m_\phi$. Therefore, the baryon number density at this time can be found as
\begin{equation}
n_B \sim \left(\frac{m_\phi}{H_{\rm inf}}\right)^2 n_{B,{\rm inf}} 
\sim \left(\frac{m_\phi}{H_{\rm inf}}\right)^{\frac{2(n-3)}{n-2}}
 \left(\frac{m_\phi}{M_P}\right)^{\frac{n}{n-2}} M_P^3,
\end{equation}
which is strongly suppressed.

\section{conclusions}
We have investigated how the Affleck-Dine baryogenesis works in the context of supersymmetric
theory. Special attention is paid to the initial condition of the Affleck-Dine field which has to
have a large VEV during and after inflation. In the usual situations, the large VEV is achieved
by a negative Hubble-induced mass term due to SUSY breaking by the finite energy density
of the inflaton. We seek for the origin of the negative Hubble-induced mass terms for
various K\"ahler potentials. 

Most important fact that we have found here is that the minima at large VEV can be obtained by
large enough Hubble-induced A terms, even if the Hubble-induced mass term is positive. Since
A terms have minima irrespective of the signature of the coupling in the nonminimal K\"ahler
potential, it is robust for the Affleck-Dine field to have large VEV during and after inflation.
Thus, the Affleck-Dine mechanism for baryogenesis works in broader classes of theories.

\section*{Acknowledgments}
The work of S.K. is supported by the Grant-in-Aid for Scientific Research from the
Ministry of Education, Science, Sports, and Culture of Japan, No.~17740156.

\appendix

\section{Hubble-induced mass and A terms}
Here we itemize the Hubble-induced mass and A terms for the minimal and nonminimal K\"ahler
potential considered above. Notice that those terms with the lowest order in $\phi$ is derived for
the Hubble-induced A terms.

\subsection{Mass terms}

$(i)$ For the minimal K\"ahler potential $K_m = \phi^\dagger \phi + I^\dagger I$,
\begin{equation}
\left[ 3 + \left( \frac{e^K|F_I|^2}{V(I)} -1 \right) \right] H^2 |\phi|^2.
\end{equation}

$(ii)$ For $K_m + \delta K_1 = K_m + \ds{\frac{a}{M_P^2}} \phi^\dagger \phi  I^\dagger I$, 
\begin{eqnarray}
& & \left\{ 3 \left[ (1-a) + (1+a)a\frac{|I|^2}{M_P^2}\right] \right.  \nonumber \\
& & +\left. \left[ (1+3a) + (1-3a)a\frac{|I|^2}{M_P^2} \right] \left( \frac{e^K|F_I|^2}{V(I)} -1 \right)
\right\} H^2 |\phi|^2. \nonumber \\
\end{eqnarray}

$(iii)$ For $K_m + \delta K_2 = K_m + \ds{\frac{b}{2M_P}} I^\dagger \phi \phi + h.c$, 
\begin{equation}
\left[ 3H^2 + b^2\frac{e^K|W_I|^2}{M_P^2} + \left( \frac{e^K|F_I|^2}{V(I)} -1 \right)H^2\right] |\phi|^2
\end{equation}

$(iv)$ For $K_m + \delta K_3 = K_m + \ds{\frac{c}{4M_P^2}} I^\dagger I^\dagger \phi \phi + h.c.$,
\begin{eqnarray}
& & \left[ 3+ \frac{3c}{2} \frac{|I|^2}{M_P^2} + \left( 1+ \frac{3c}{2}  \frac{|I|^2}{M_P^2} 
- \frac{c^2}{4} \frac{|I|^4}{M_P^4} \right) \left( \frac{e^K|F_I|^2}{V(I)} -1 \right) \right] \nonumber \\
& & \hspace{55mm} \times H^2 |\phi|^2.
\end{eqnarray}

$(v)$ For $K_m + \delta K_4 = K_m + \ds{\frac{d}{M_P}} I \phi^\dagger \phi + h.c. $
\begin{eqnarray}
\left\{ 3 \left[ 1 + d \frac{I+I^\dagger}{M_P} 
+ d^2 \left( 1 + d \frac{I+I^\dagger}{M_P} \right)^{-1} \right] \right. \quad\quad & & \nonumber \\
+ \left. \left[ 1 + d \frac{I+I^\dagger}{M_P} 
+ 3 d^2 \left( 1 + d \frac{I+I^\dagger}{M_P} \right)^{-1} \right] \right. & & \nonumber \\ 
\hspace{20mm} \times \left. \left( \frac{e^K|F_I|^2}{V(I)} -1 \right) \right\} H^2 |\phi|^2. & & 
\end{eqnarray}

\subsection{A-terms}
$(i)$ For the minimal K\"ahler potential $K_m = \phi^\dagger \phi + I^\dagger I$,
\begin{equation}
W_\phi \,\phi \,\frac{e^K W^*(I^\dagger)}{M_P^2} 
+ W(\phi) \frac{I^\dagger}{M_P}\frac{e^K F_{\bar{I}}^*}{M_P}
- 3 W(\phi) \frac{e^K W^*(I^\dagger)}{M_P^2} + h.c.
\end{equation}

$(ii)$ For $\delta K_1 = \ds{\frac{a}{M_P^2}} \phi^\dagger \phi  I^\dagger I$, 
\begin{eqnarray}
a W_\phi \, \phi \frac{|I|^2}{M_P^2}\frac{e^K W^*(I^\dagger)}{M_P^2}  
-a  W(\phi) \frac{I^\dagger}{M_P} \frac{|I|^2}{M_P^2}  \frac{e^K F^*_{\bar{I}}}{M_P}
& & \nonumber \\
\qquad - a W_\phi \, \phi \,  \frac{I^\dagger}{M_P}   \left( 1 - a \frac{|I|^2}{M_P^2} \right) 
\frac{e^K F_{\bar{I}}^*}{M_P} + h.c.  & & \\
\nonumber
\end{eqnarray}

$(iii)$ For $\delta K_2 = \ds{\frac{b}{2M_P}} I^\dagger \phi \phi + h.c$, 
\begin{equation}
b W_\phi \, \phi^\dagger  \frac{I}{M_P}  \frac{e^K W^*(I^\dagger)}{M_P^2}  
- b W_\phi \, \phi^\dagger  \frac{e^K F_{\bar{I}}^*}{M_P} + h.c.
\end{equation}
In addition to these, there are terms of another type as
\begin{equation}
3bH^2 \frac{I^\dagger}{M_P} \phi \phi -\frac{b}{2} \frac{e^K W^*(I^\dagger)}{M_P^2}  
\left( \frac{W_I}{M_P} - \frac{I^\dagger}{M_P} \frac{W(I)}{M_P^2} \right) \phi \phi + h.c.
\end{equation}

$(iv)$ For $\delta K_3 =  \ds{\frac{c}{4M_P^2}} I^\dagger I^\dagger \phi \phi + h.c.$,
\begin{equation}
\frac{c}{2} W_\phi \, \phi^\dagger  \frac{II}{M_P^2} \frac{e^K W^*(I^\dagger)}{M_P^2} 
- c W_\phi \, \phi^\dagger  \frac{I}{M_P} \frac{e^K F_{\bar{I}}^*}{M_P} + h.c.
\end{equation}
In addition to these, there are terms of another type as
\begin{equation}
\frac{3c}{4} H^2 \frac{I^\dagger I^\dagger}{M_P^2} \phi \phi
- \frac{c}{2} \frac{I^\dagger}{M_P}\frac{e^KW^*(I^\dagger)}{M_P^2} \frac{W_I}{M_P} \phi\phi+ h.c.
\end{equation}

$(v)$ For $\delta K_4 = \ds{\frac{d}{M_P}} I \phi^\dagger \phi + h.c.$,
\begin{equation}
- d \, W(\phi) \frac{e^K F_{\bar{I}}^*}{M_P} + h.c.
\end{equation}




\begin{thebibliography}{90}

\bibitem{BBN}
For recent review, see e.g., 
J.~P.~Kneller and G.~Steigman,
New J.\ Phys.\  {\bf 6}, 117 (2004).


\bibitem{WMAP}
D.~N.~Spergel {\it et al.},
arXiv:astro-ph/0603449.


\bibitem{AD}
I.~Affleck and M.~Dine,
Nucl.\ Phys.\ B {\bf 249}, 361 (1985).

\bibitem{DRT}
M.~Dine, L.~Randall and S.~Thomas,
Nucl.\ Phys.\ B {\bf 458}, 291 (1996).


\bibitem{qball}
A.~Kusenko and M.~E.~Shaposhnikov,
Phys.\ Lett.\ B {\bf 418}, 46 (1998);
K.~Enqvist and J.~McDonald,
Nucl.\ Phys.\ B {\bf 538}, 321 (1999);
S.~Kasuya and M.~Kawasaki,
Phys.\ Rev.\ D {\bf 61}, 041301(R) (2000);
S.~Kasuya and M.~Kawasaki,
Phys.\ Rev.\ D {\bf 62}, 023512 (2000);
S.~Kasuya and M.~Kawasaki,
Phys.\ Rev.\ Lett.\  {\bf 85}, 2677 (2000);
S.~Kasuya and M.~Kawasaki,
Phys.\ Rev.\ D {\bf 64}, 123515 (2001);
S.~Kasuya, M.~Kawasaki and F.~Takahashi,
Phys.\ Rev.\ D {\bf 68}, 023501 (2003).

\bibitem{McDonald}
J.~McDonald,
Phys.\ Lett.\ B {\bf 456}, 118 (1999).

\bibitem{Linde} A.~Linde,
  {\it Particle Physics and Inflationary Cosmology} (Harwood, Chur,
  Switzerland, 1990).

\end{thebibliography}
\end{document}